\documentclass[a4paper, 10pt, twocolumn]{article}

\usepackage{amsfonts,amssymb,amsmath,amsthm}
\usepackage{subfigure}
\usepackage{graphicx}
\usepackage[footnotesize]{caption}
\usepackage{xcolor}
\usepackage{import}
\usepackage{verbatim}
\usepackage{dsfont}

\usepackage[top=1.5cm, left=1.5cm, right=1.5cm, bottom=1.5cm]{geometry}

\title{Phase retrieval with Bregman divergences: \\ Application to audio signal recovery}

\author{Pierre-Hugo Vial$^1$, Paul Magron$^1$, Thomas Oberlin$^{1,2}$ and Cédric Févotte$^1$\thanks{This work is supported by the European Research Council (ERC FACTORY-CoG-6681839).}\\
\footnotesize $^1$IRIT, Université de Toulouse, CNRS, Toulouse, France\ $^2$ISAE-SUPAERO, Toulouse, France.
}
\date{\empty} 

\renewenvironment{abstract}{\bf\small {\em\ Abstract---}}{}

\begin{document}

\maketitle

\begin{abstract} Phase retrieval aims to recover a signal from magnitude or power spectra measurements. It is often addressed by considering a minimization problem involving a quadratic cost function. We propose a different formulation based on Bregman divergences, which encompass divergences that are appropriate for audio signal processing applications. We derive a fast gradient algorithm to solve this problem.
\end{abstract}

\section{Introduction}
\label{sec:introduction}

Reconstructing data from phaseless measurements is at the core of many signal processing problems in various domains such as X-ray crystallography \cite{Harrison93}, optics \cite{Walther63} and astronomy \cite{Fienup87}. The reconstruction task is usually treated as an optimization problem involving a quadratic cost function, which is minimized by, e.g., gradient descent \cite{WF}, alternating projections \cite{GS}, majorization-minimization \cite{PRIME}, and alternating direction method of multipliers \cite{Liang18, Wen12}.

In particular, many audio signal processing applications such as speech enhancement or source separation operate on the magnitude of the short-time Fourier transform (STFT), also called spectrogram. Recovering the missing phase from a modified spectrogram is then required in order to synthesize time-domain signals.
The Griffin-Lim algorithm (GLA) \cite{GLA}, a version of the alternating projections algorithm proposed by Gerchberg and Saxton \cite{GS} adapted to STFT measurements, is one of the most popular techniques in the literature and can be considered as a baseline for audio signal recovery. Improvements of GLA include an accelerated version (Fast GLA) \cite{FGLA} and real-time purposed versions \cite{RTISI, RTISILA}.

In this work, we propose to replace the quadratic loss in the phase retrieval (PR) optimization problem by a general Bregman divergence. It encompasses divergences such as $\beta$-divergences \cite{betadiv}, with special cases such as Kullback-Leibler and Itakura-Saito that are acknowledged for enabling better performance in audio spectral decomposition \cite{Fevotte09,Smaragdis2014}.
The resulting objective is minimized by accelerated gradient descent, with an efficient computation of the gradient involving STFT and inverse STFT. We illustrate the potential of this new approach for some well-chosen divergences on speech signals.

\section{Phase retrieval}
\label{sec:first-section}

\par \textbf{Phase retrieval.} PR consists in reconstructing a signal $\mathbf x_0 \in \mathbb C^L$ from nonnegative measurements $\mathbf r \approx |\mathbf A \mathbf{x}_0|^d \in \mathbb R^K$, where the power is applied element-wise and $\mathbf A \in \mathbb C^{K\times L}$ is the measurement matrix. Usually, $\mathbf r$ corresponds to magnitude ($d=1$) or power ($d=2$) measurements. PR is often formulated as the following optimization problem:
\begin{equation}
    \label{eq:pr}
    \underset{\mathbf x \in \mathbb R^L}{\text{min}}\quad \|\mathbf r - |\mathbf A \mathbf x|^d\|_2 .
\end{equation}

\noindent \textbf{Short-time Fourier transform.} In this work, the measurement operator is the STFT. For every signal $\mathbf x \in \mathbb R^L$, the STFT coefficient at time $n\in \{0,\dots,N-1\}$ and frequency $m\in \{0,\dots,M-1\}$ is the dot product
\begin{equation}
    (\mathbf A \mathbf x) [m+nM] = \langle \mathbf x , \mathbf a^*_{m+nM}\rangle ,
\end{equation}
\begin{equation} \nonumber
    \mbox{where } \mathbf a_{m+nM}[l]=w_a(l-nH)e^{-2i\pi\frac{m}{M}l}\quad\forall l\in \{0,\dots,L-1\},
\end{equation}
with $w_a$ the analysis window and $H$ the hop size.
The inverse STFT operator $\mathbf B$ is defined for every sequence of complex coefficients $\tilde{\mathbf y} \in \mathbb C^{MN}$ as
\begin{equation}
    (\mathbf B \tilde{\mathbf y})[l]=\sum_{m,n} \tilde{\mathbf y}[m+nM]\mathbf b_{m+nM}^*[l],
\end{equation}
\begin{equation} \nonumber
    \mbox{where }\mathbf b_{m+nM}[l]= w_s(l-nH)e^{-2i\pi\frac{m}{M}l}\quad\forall l\in \{0,\dots,L-1\}.
\end{equation}
Under duality assumptions for $w_a$, $w_s$ and $H$, one can show that $\mathbf B = \mathbf A^\mathsf{H}$  \cite{Portnoff80}. In the following, the vectorized time-frequency coefficients are denoted with a tilde, i.e., $\tilde{\mathbf x} = \mathbf A \mathbf x$.

\noindent \textbf{Griffin-Lim algorithm.} GLA \cite{GLA} solves (\ref{eq:pr}) in the STFT domain by alternating projections on $\mathcal M$, the set of coefficients that satisfies the magnitude constraints, and $\mathcal C$, the set of consistent coefficients:
\begin{equation}
    \mathcal P_\mathcal M(\tilde{\mathbf x})=\mathbf r \odot \frac{\tilde{\mathbf x}}{|\tilde{\mathbf x}|} \mbox{ and }
    \mathcal P_\mathcal C(\tilde{\mathbf x})=\mathbf A \mathbf A^\mathsf{H}\tilde{\mathbf x},
\end{equation}
where $\odot$ denotes the element-wise product. GLA is proved to converge to a critical point of the measure minimized in (\ref{eq:pr}) when $d=1$. Like the Gerchberg-Saxton algorithm (GSA) \cite{GS}, GLA is an alternating projection algorithm. They however differ as GSA considers the DFT as the measurement matrix and adds a supplementary constraint on the signal to make the solution unique. In GLA, this constraint is not necessary as uniqueness is brought by the redundancy of STFT.

\section{Algorithm}
\label{sec:second-section}

\par We propose a generalization of (\ref{eq:pr}) to the family of Bregman divergences. The problem writes
\begin{equation}
\label{eq:prb}
    \underset{\mathbf x\in \mathbb R^L}{\text{min}}\quad J( \mathbf x) :=\mathcal D_ \psi(\,\mathbf r\,|\,|\mathbf A\mathbf x|^d\,),
\end{equation}
where $\mathcal D_\psi$ is the Bregman divergence associated to $\psi$ :
\begin{equation}
    \mathcal  D_\psi(\,\mathbf x\,|\,\mathbf y\,)=\psi(\mathbf x) - \psi(\mathbf y) - \langle\nabla\psi(\mathbf y), \mathbf x-\mathbf y \rangle.
\end{equation}
As illustrated by the Table \ref{tab:bregdiv}, many divergences or distances can be expressed as a Bregman divergence.

\begin{table}[h]
    \centering
    \begin{tabular}{|c|c|} \hline
        Divergence & Generating function $\psi(\mathbf x)$\\ \hline
        $\ell_2$ &  $\|\mathbf x\|_2^2$ \\ \hline
        Kullback-Leibler & $\mathbf x \log \mathbf x$ \\ \hline
        Itakura-Saito & $- \log \mathbf x$ \\ \hline
        $\beta$-divergence & $\frac{\mathbf x^\beta }{\beta(\beta-1)}-\frac{\mathbf x}{\beta -1}+\frac{1}{\beta}$\\ \hline
    \end{tabular}
    \caption{Examples of usual Bregman divergences}
    \label{tab:bregdiv}
\end{table}

Similarly to \cite{WF}, we propose a gradient algorithm to minimize the criterion in (\ref{eq:prb}). As the latter is not holomorphic, we use the Wirtinger formalism in order to express the gradient, which yields: 
\begin{align}
    \nabla J( \mathbf x) = \mathbf A^\mathsf{H} \Big \{ &(\mathbf A \mathbf x) \odot \frac{d}{2}|\mathbf A \mathbf x|^{d-2} \\ 
    \nonumber&\odot \left (\nabla^2\psi(|\mathbf A \mathbf x|^d)^\mathsf{T}(|\mathbf A \mathbf x|^d-\mathbf r)\right)^* \Big \}.
\end{align}
We consider a simple gradient algorithm with a constant step in order to compare the reconstruction performances while minimizing (\ref{eq:prb}) with different divergences. One can note that in the particular case $\psi = \|\cdot\|_2^2$, \ (\ref{eq:pr}) and (\ref{eq:prb}) are equivalent. Moreover, when $d=1$, the gradient writes 
\begin{equation}
    \nabla J(\mathbf x) = \mathbf x - \mathbf A^\mathsf{H} \Big \{ \mathbf r \odot \frac{\mathbf A \mathbf x}{|\mathbf A \mathbf x|}\Big \}.
\end{equation}
Thus, when $\psi = \|\cdot\|_2^2$, the gradient algorithm with a step size equal to $1$ is equivalent to GLA.

\section{Experiments}
\label{sec:third-section}

\textbf{Protocol}. We perform experiments on $10$ utterances of the LibriSpeech \cite{LibriSpeech} speech database. All are monochannel, cropped to 2-seconds duration and sampled at $22050$ Hz. 
The STFT uses 512 samples-long Hann windows and 50 \% overlap.
As phase retrieval is usually performed on modified measurements for many audio tasks, we simulate that situation by adding a Gaussian white noise at various signal-to-noise ratios (SNR) to the temporal signal. We use spectral subtraction \cite{Boll79} in the STFT domain to produce the nonnegative measurements $\mathbf r$.

Our gradient algorithm uses a constant step size equal to 1. As in Fast GLA \cite{FGLA}, a Nesterov-like acceleration scheme is used, with an acceleration parameter equal to $0.99$. Several divergences were considered, among which we present the following, as they yielded the best performances. We propose $\ell_2$ distance and $\beta$-divergence with $\beta=0.5$ for the magnitude problem ($d=1$), and Kullback-Leibler divergence for the power problem ($d=2$). 
Finally, we also studied a variant of the latter, which consists in minimizing $\mathcal D_{KL}(|\mathbf A \mathbf x|^d \, |\, \mathbf r)$, thus exploiting the non-symmetry of the divergence (which we do not detail here for brevity). The resulting algorithm is denoted ``KL left" and uses a step size equal to $0.4$.
As comparison baselines, we also test GLA and Fast GLA. All algorithms use $1000$ iterations.

\noindent \textbf{Performance criteria}. The reconstruction quality is evaluated with the spectral convergence metric defined as
\begin{equation}
\label{eq:sc}
    E_{SC}(\mathbf r, \mathbf x) = \frac{\|\ \mathbf r^{1/d} - |\mathbf A\mathbf x|\|_2}{\|\mathbf r^{1/d}\|_2}.
\end{equation}
We also consider the short-term objective intelligibility (STOI) measure \cite{STOI} in order to compare the performance of the different algorithms on the speech signal from a perceptual point of view.

\begin{figure}[t]
    \centering
    \includegraphics[scale=0.55]{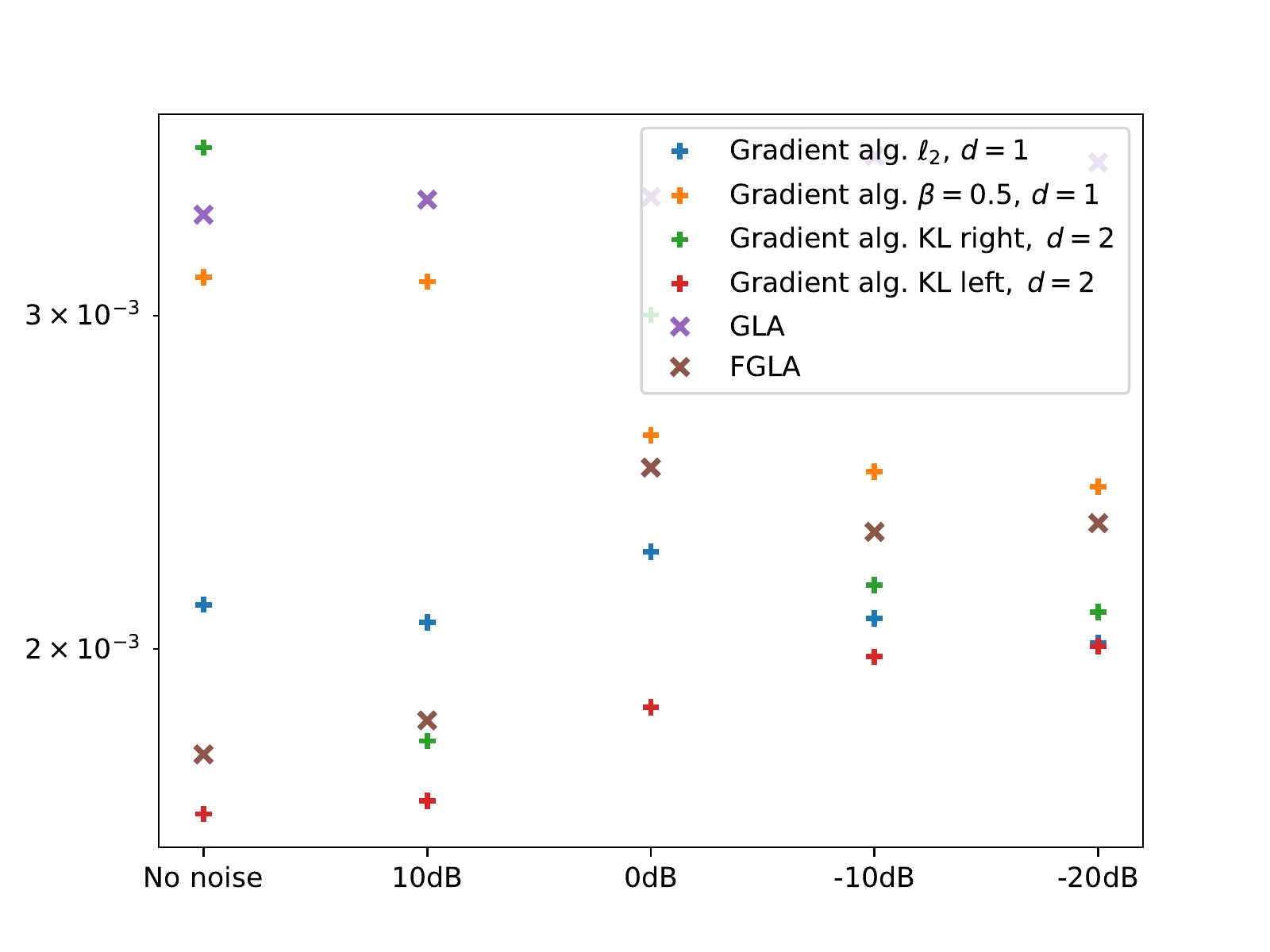}
    \caption{Mean spectral convergence values for various input SNR. A lower value corresponds to a better spectrogram approximation.}
    \label{fig:fig1}
\end{figure}

\noindent \textbf{Results}. Figure \ref{fig:fig1} presents the mean results in terms of spectral convergence for several levels of input SNR. The proposed gradient algorithms outperform GLA and Fast GLA in terms of spectral convergence in some situations. In particular, the ``KL left'' formulation seems more robust to the input perturbation.

\begin{figure}[t]
    \centering
    \includegraphics[scale=0.55]{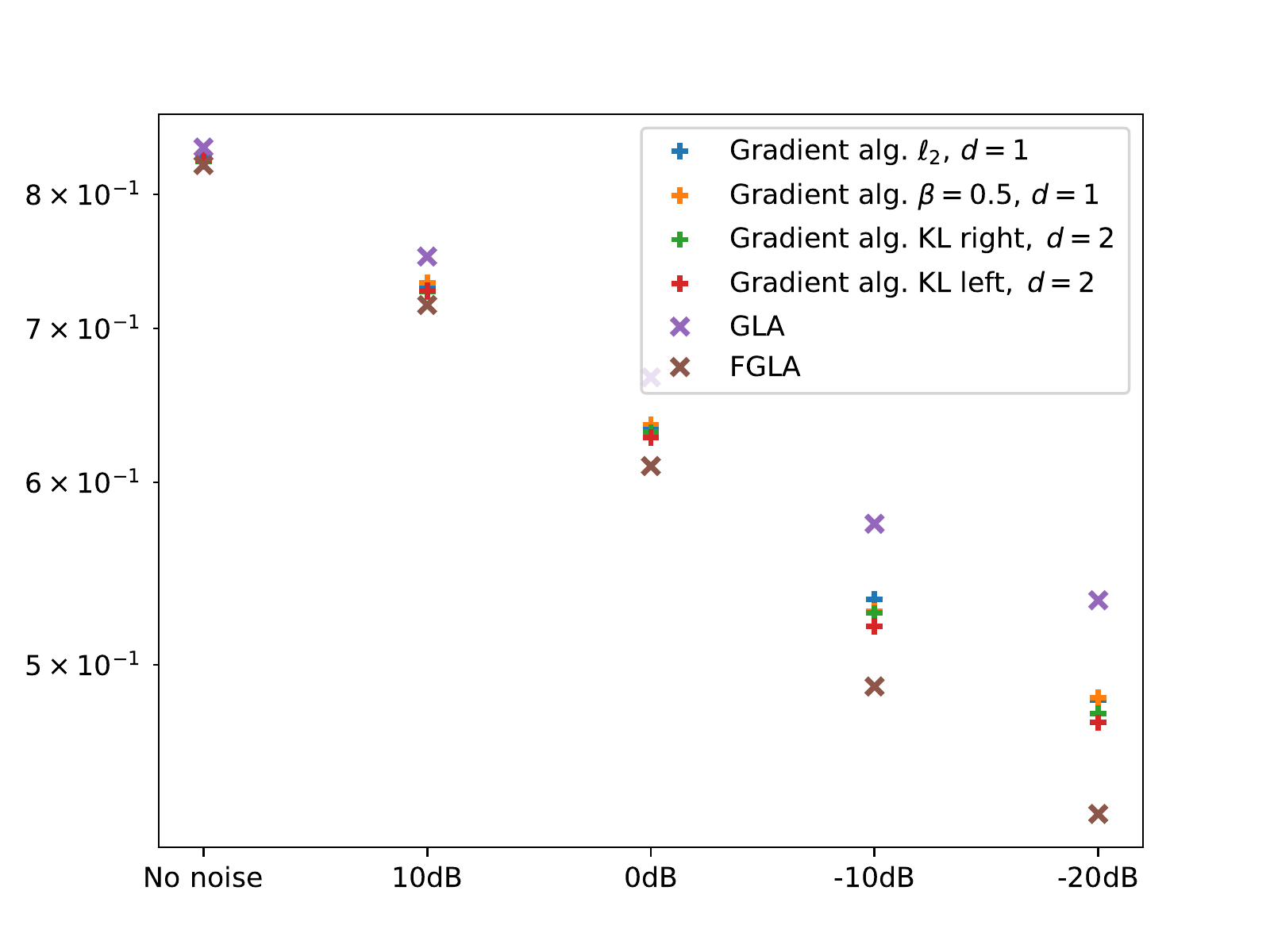}
    \caption{Mean STOI values for various input SNR. STOI ranges from $0$ to $1$, a higher value corresponds to a better intelligibility.}
    \label{fig:fig2}
\end{figure}

Figure \ref{fig:fig2} shows the STOI score obtained with the different algorithms. All algorithms show similar intelligibility performance when no noise is added. When the SNR lowers, the intelligibility decreases. Surprisingly, GLA is the less affected by this phenomenon while its performances in terms of spectral convergence are among the weakest. Finding a good performance measure both in terms of spectral approximation and perceptual audio quality still seems to be an open question and makes the interpretation of the observed results a complex task.

\section{Conclusion}
\label{sec:conclusion}
We have addressed the problem of phase retrieval with the Bregman divergences in the context of audio signal recovery. We derived a gradient algorithm and made an explicit link with GLA when the cost function is quadratic. In future work, we will consider more advanced gradient schemes with adaptive step size and extend our study to musical signals.

\clearpage


\end{document}